\documentclass[a4paper]{jpconf}

\usepackage{graphicx}

\begin{document}

\title{Connecting $B_d$ and $B_s$ decays through QCD factorisation and flavour
symmetries}

\author{S\'ebastien Descotes-Genon}

\address{Laboratoire de Physique Th\'eorique,\\ CNRS/Univ. Paris-Sud 11 (UMR 8627),
91405 Orsay Cedex, France}

\ead{descotes@th.u-psud.fr}

\begin{abstract}
We analyse $B_{d,s}\to K^{(*)0}K^{(*)0}$ modes within the SM, 
relating them in a controlled way through $SU(3)$-flavour symmetry 
and QCD-improved factorisation. We propose a set of sum rules for 
such penguin-mediated decays to constrain some CKM angles. We determine $B_s\to KK$
branching ratios and CP-asymmetries as functions of 
$A_{dir}(B_d \to K^0{\bar K}^0)$. Applying the same techniques to
$B_{d,s}\to K^{*0}K^{*0}$, we outline strategies to determine
the $B_s$ mixing angle.
\end{abstract}

Non-leptonic two-body $B_d$- and $B_s$-decays provide many interesting
ways of testing the CKM mechanism of CP-violation, but
the effects of strong interaction often hinder quantitative predictions.
The relevant hadronic quantities can be estimated through
flavour symmetries, such as $U$-spin, but with a sizeable uncertainty. 
QCD factorisation (QCDF) provides a complementary tool, specially
for short distances, but this
expansion in $\alpha_s$ and $1/m_b$ cannot predict 
some $1/m_b$-suppressed long-distance effects.
Recently, it was proposed to improve theoretical predictions
by combining QCDF and $U$-spin in particular classes 
decays~\cite{DMV1,DMV2}.

\section{Sum rules}

In the Standard Model (SM), we can always split a $B$-decay amplitude into its
tree and penguin contributions $
\bar{A}\equiv A(\bar{B}_q\to M \bar{M})
  =\lambda_u^{(q)} T_M^{q} + \lambda_c^{(q)} P_M^{q}$
according to the CKM factors $\lambda_p^{(q)}=V_{pb}V^*_{pq}$.
One can compute these contributions for 
$\bar{B}_s\to K^0 \bar K^0$ within QCDF~\cite{BBNS,BN}:
\begin{eqnarray}
\label{eq3}
{\hat T^{s\, 0}} &=&
  \bar\alpha_4^u-\frac{1}{2}\bar\alpha_{4EW}^u
    +\bar\beta_3^u +2 \bar\beta_4^u - \frac{1}{2} \bar\beta^u_{3EW} -
  \bar\beta^u_{4EW} \,,
\\  \label{eq4}
{\hat P^{s\, 0}} &=&
 \bar\alpha_4^c-\frac{1}{2}\bar\alpha_{4EW}^c
    +\bar\beta_3^c +2 \bar\beta_4^c
    - \frac{1}{2} \bar\beta^c_{3EW} - \bar\beta^c_{4EW} \,,
\end{eqnarray}
where $\hat P^{sC}=P^{sC}/A^s_{KK}$, $\hat T^{sC}=T^{sC}/A^s_{KK}$
and $A^q_{KK}=M^2_{B_q} F_0^{\bar{B}_q\to K}(0) f_K
{G_F}/{\sqrt{2}}$. $\beta$'s denote weak-annihilation contributions whereas
$\alpha$'s collect remaining terms (vertex and hard-spectator interactions).
A similar structure occurs fro 
the tree and penguin contributions ${T^{d\, 0}}$ and $P^{d\, 0}$ for
$\bar{B}_d\to K^0 \bar K^0$, and for longitudinally polarised 
$K^{*0}\bar{K}^{*0}$~\cite{BN,DMV2}.
As exemplified in eqs.~(\ref{eq3})-(\ref{eq4}), 
for penguin-mediated decays, $T$ and $P$ are actually generated only
by penguin topologies, and thus share the same
long-distance dynamics: the difference comes from the ($u$ or $c$)
quark running in the loops~\cite{DMV1}. Thus, $\Delta=T-P$ is midly
affected by annihilation and hard-spectator contributions, and it can be
computed with smaller uncertainties than $T$ or $P$ individually within QCDF: 
$
\Delta^{d0}\equiv T^{d0}-P^{d0}=A_{KK}^d[\alpha_4^{u}-\alpha_4^{c} +
\beta_3^{u}-\beta_3^{c}+2\beta_4^{u}-2\beta_4^{c}]$.
The $1/m_b$ suppressed long-distance dynamics, modelled in QCDF,
cancels in the differences between $u$ and $c$ contributions.

These theoretically well-behaved differences
 are related to the CP-averaged branching ratio $BR$
and the direct and mixed CP-asymmetries $\mathcal{A}_{\rm dir}$ and 
$\mathcal{A}_{\rm mix}$ (see~\cite{DMV1,DMV2} for the exact definitions).
For a $B_d$ meson decaying through a $b\to D$
process $(D=d,s)$ [such as $B_{d}\to K^{*0} \bar{K}^{*0}$
or $B_{d}\to \phi
  \bar{K}^{*0}$ (with a subsequent decay into a CP eigenstate)], 
one extracts $\alpha$ \cite{DILM}
and $\beta$ from:
\begin{eqnarray}
\sin^2{\alpha}&=&{\widetilde{BR}}/({2|\lambda_u^{(D)}|^2|\Delta|^2})\left( 1-
\sqrt{1-{(\mathcal{A}_{\rm dir})}^2-{(\mathcal{A}_{\rm mix})}^2} \right)\,,\\
\sin^2{\beta}&=&{\widetilde{BR}}/({2|\lambda_c^{(D)}|^2|\Delta|^2})\left( 1-\sqrt{1-
{(\mathcal{A}_{\rm dir})}^2-{(\mathcal{A}_{\rm mix})}^2} \right)\,,
\end{eqnarray}
where $\widetilde{BR}$ is the CP-averaged branching ratio, up to a trivial kinematic 
factor~\cite{DMV2}.
Similar identities can be used for a $B_s$ meson 
decaying through $b\to D$ $(D=d,s)$ [such as $B_{s}\to K^{*0} \bar{K}^{*0}$
or $B_{s}\to \phi \bar{K}^{*0}$]
to extract the angles $\beta_s$
and $\gamma$, assuming no New Physics in the decay.


\section{$B_d\to K^0\bar{K}^0$ and $B_s\to K^0\bar{K}^0$}

These penguin-mediated decays are related by
$U$-spin, with a small breaking: few processes
(weak annhilation and spectator interaction) probe the
spectator quark,
as confirmed by QCDF:
\begin{eqnarray} 
P^{s0}&=&f P^{d0}
  \Big[1+(A^d_{KK}/P^{d0})\Big\{\delta\alpha_4^c
           -\delta\alpha_{4EW}^c/2
  +\delta\beta_3^c+2\delta\beta_4^c
  -\delta\beta_{3EW}^c/2
  -\delta\beta_{4EW}^c
    \Big\}\Big]\,,\\
T^{s0}&=&f T^{d0}
  \Big[1+(A^d_{KK}/T^{d0})\Big\{
  \delta\alpha_4^u-\delta\alpha_{4EW}^u/2
   +\delta\beta_3^u
   +2\delta\beta_4^u
   -\delta\beta_{3EW}^u/2
   -\delta\beta_{4EW}^u
    \Big\}\Big]\,,
\end{eqnarray}
These ratios involve the $U$-spin breaking differences
$\delta\alpha_i^p\equiv\bar\alpha_i^p-\alpha_i^p$ (id. for $\beta$).
Apart from the ratio 
$
f={M_{B_s}^2 F_0^{\bar{B}_s\to K}(0)}/{[M_{B_d}^2 F_0^{\bar{B}_d\to K}(0)]}
$,
$U$-spin arises only through $1/m_b$-suppressed
terms in which most long-distance effects have cancelled out.
In agreement with this observation, QCDF~\cite{BN} yields tiny uncertainties:
$|P^{s0}/(fP^{d0})-1| \leq 3 \%$ and $|T^{s0}/(fT^{d0})-1| \leq 3 \%$.
These relations depend much less on the QCDF model for $1/m_b$-suppressed
contributions than the predictions for indivdiual tree or penguin
contributions, and thus they provide an interesting alternative to a pure 
QCDF computation. One can also relate the penguin contributions to 
$\bar{B}_d \to K_0\bar{K}_0$ and
$\bar{B}_s \to K^+ K^-$ (see~\cite{DMV1} for the treatment of 
tree contributions).

For $B_d\to K^0\bar{K}^0$, the branching ratio
$BR^{d0}=(0.96\pm 0.25)\cdot 10^{-6}$~\cite{brd} has been measured.
If $A_{dir}^{d0}$ becomes available,
we may exploit the theoretically well-controlled value of 
$\Delta_d\equiv T^{d0}-P^{d0}$
to get the two moduli and the relative phase
of $T^{d0}$ and $P^{d0}$ from $BR^{d0}$,
$A_{dir}^{d0}$ and $\Delta_d$. Then we can use the previous bounds to
compute the tree and penguin contributions for $B_s\to KK$ decays, leading to 
the SM predictions for the corresponding observables (see Table I in
ref.~\cite{DMV1}):
$Br(B_s\to K^0 \bar{K}^0) = (18\pm 7\pm 4\pm 2)\cdot 10^{-6}$ and
$Br(B_s\to K^+ \bar{K}^-) = (20\pm 8\pm 4\pm 2)\cdot 10^{-6}$, the latter
being in very good agreement with the latest CDF measurement~\cite{cdf}. 

\section{$B_d\to K^{*0}\bar{K}^{*0}$ and $B_s\to K^{*0}\bar{K}^{*0}$}

\begin{figure}
\begin{center}
\includegraphics[width=7.5cm]{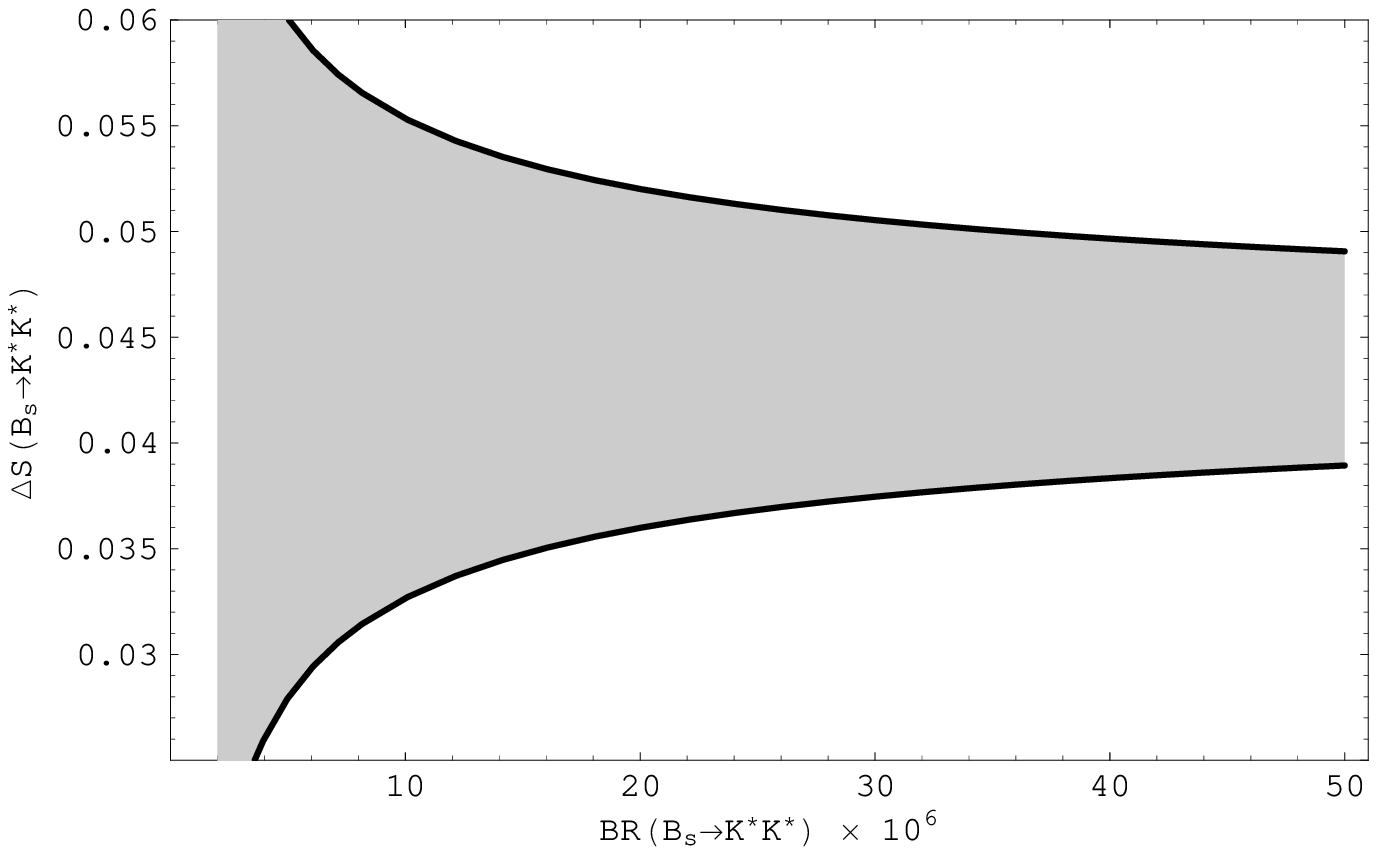}
\includegraphics[width=7.5cm]{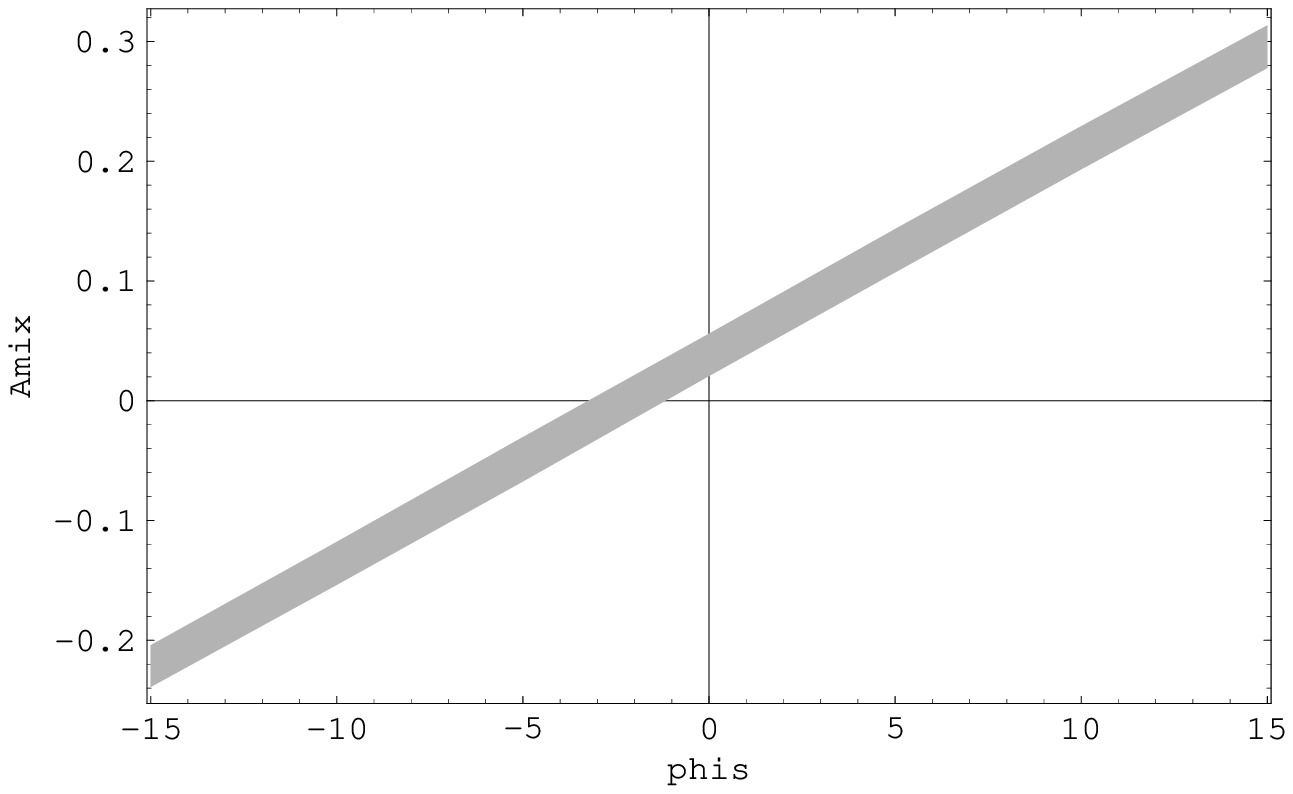}
\end{center}
\caption{The absolute bounds on $\Delta S(B_s\to K^{*0} \bar{K}^{*0})$
as functions of $BR^{\rm long}(B_s\to K^{*0} \bar{K}^{*0})$ (on the left) and
the relation between $\mathcal{A}_{\rm mix}^{\rm long}(B_s\to K^{*0}\bar{K}^{*0})$
and the $B_s-\bar{B}_s$ mixing angle $\phi_s$ (on the right), assuming
$BR^{\rm long}(B_d\to K^{*0}\bar{K}^{*0})\geq 5 \times 10^{-7}$ and
$\gamma=62^\circ$.} 
\label{fig:DeltaS}
\end{figure}

We focus on observables for mesons with a longitudinal
polarisation which can be measured experimentally and predicted theoretically
with a good accuracy. $B_s\to K^{*0} \bar{K}^{*0}$ is
in principle a clean mode to extract the mixing angle $\phi_s$. 
An expansion in powers of $\lambda_u^{(s)}/\lambda_c^{(s)}$ yields
\begin{equation}
\label{pr}\mathcal{A}_{\rm mix}^{\rm long}(B_s\to K^{*0} \bar{K}^{*0})\simeq \sin{\phi_s}+2\left|
\frac{\lambda_u^{(s)}}{\lambda_c^{(s)}} \right|
{\rm Re}\left( \frac{T^{s0*}}{P^{s0*}} \right) \sin{\gamma}\cos{\phi_s}+\cdots
 =\sin{\phi_s}+\Delta S(B_s\to K^{*0} \bar{K}^{*0})
\end{equation}
A significant value of
$T^{s0*}/P^{s0*}$  
could spoil the extraction of $\sin{\phi_s}$. One can use our knowledge on 
$T^{s0*}-P^{s0*}$ to bound 
$\Delta S(B_s\to K^{*0} \bar{K}^{*0})$, as illustrated in fig.~\ref{fig:DeltaS}.

One can also relate the observables in $B_{d,s}\to K^{*0} \bar{K}^{*0}$ through
the same combination of $U$-spin symmetry and QCDF. Once again, $U$-spin is
mainly broken through the ratio of relevant form factors $f^*$, whereas most of 
the long-distance annihilation and spectator scattering contributions cancel
in $P^{s0*}/(f^* P^{d0*})$ and
$T^{s0*}/(f^* T^{d0*})$. Indeed, QCDF yields
$|P^{s0*}/(f^* P^{d0*})-1|\le 12\%$
and $|T^{s0*}/(f^* T^{d0*})-1|\le 15\%$, which
can be exploited to predict $B_s\to K^{*0} \bar{K}^{*0}$ observables.
The ratio of branching ratios $BR^{\rm long}(B_s\to K^{*0}\bar{K}^{*0})/BR^{\rm long}(B_d\to
K^{*0}\bar{K}^{*0})$ and the asymmetries as predicted in the SM
turn out to be quite insensitive to
the exact value of  $BR^{\rm long}(B_d\to K^{*0}\bar{K}^{*0})$ as long as
$BR^{\rm long}(B_d\to K^{*0}\bar{K}^{*0})\geq 5 \times 10^{-7}$:
\begin{eqnarray}
&& {BR^{\rm long}(B_s\to K^{*0}\bar{K}^{*0})}/{BR^{\rm long}(B_d\to K^{*0}\bar{K}^{*0})}=17\pm 6 \\ 
&&\mathcal{A}_{\rm dir}^{\rm long}(B_s\to K^{*0}\bar{K}^{*0})=0.000\pm 0.014 
\qquad \mathcal{A}_{\rm mix}^{\rm long}(B_s\to K^{*0}\bar{K}^{*0})=0.004\pm 0.018
\end{eqnarray}
If one assumes no New Physics in the decay $B_s\to
K^{*0}\bar{K}^{*0}$, this method relates directly $\mathcal{A}_{\rm mix}^{\rm long}(B_s\to
K^{*0}\bar{K}^{*0})$ and $\phi_s$ as indicated in fig.\ref{fig:DeltaS}.

\section{Conclusions} 

We have combined experimental data, flavour
symmetries and QCDF to gain control on 
penguin-mediated $B_{d,s}$ decays. 
The difference between tree and penguin contributions
can be assessed with a good accuracy.
The $U$-spin breaking between $B_d$ and $B_s$ modes 
arises in few factorisable corrections 
(ratio of form factors)
and non-factorisable corrections (weak annihilation and 
spectator scattering). QCDF confirms these expectations, and provides
predictions with a limited model dependence on $1/m_b$-suppressed
long-distance contributions. We outlined the implications for $B_s\to K\bar{K}$
in pseudoscalar and vector channels.
Sizeable NP effects would break these SM correlations between $B_d$
and $B_s$ decays, leading to
departure from our predictions. 

\ack
Work supported in part by EU Contract No. MRTN-CT-2006-035482, \lq\lq FLAVIAnet''.

\end{document}